\documentclass[prd,reprint,eqsecnum,floats,aps,amsmath,amssymb,nofootinbib,superscriptaddress,showpacs,longbibliography]{revtex4-1}

\usepackage[utf8]{inputenc}
\usepackage[T1]{fontenc}
\usepackage[english]{babel}

\usepackage{amsmath,amsthm,amssymb,amsfonts}
\usepackage{xfrac}
\usepackage{tensor}
\usepackage{braket}
\usepackage{mathtools}
\usepackage{graphicx}
\usepackage{subfigure}

\usepackage{bbold}

\usepackage{lmodern}
\usepackage{microtype}

\numberwithin{equation}{section}
\numberwithin{thr}{section}
\numberwithin{chr}{section}
\numberwithin{df}{section}

\renewcommand{\(}{\left(}
\renewcommand{\)}{\right)}

\newcommand{\pd}[2]{\frac{\partial #1}{\partial #2}}

\renewcommand{\phi}{\varphi}
\newcommand{\ju}{j^u}
\newcommand{\jd}{j^d}
\newcommand{\jud}{j^{u+d}}

\newcommand{\AUA}{\alpha \( U, \, A \)}

\newcommand{\twoD}{{}^{(2)} \hspace{-0.5ex}}

\DeclareMathOperator{\ent}{\mathcal{S}}
\DeclareMathOperator{\arcosh}{arcosh}
\begin{document}

\title{The geometry and entanglement entropy of surfaces in loop quantum gravity}

\author{David Gr\"uber}
\email{david.grueber@fau.de}
\author{Hanno Sahlmann}
\email{hanno.sahlmann@gravity.fau.de}
\author{Thomas Zilker}
\email{thomas.zilker@gravity.fau.de}
\affiliation{Institute for Quantum Gravity\\ Friedrich-Alexander-Universität Erlangen-Nürnberg\\ Staudtstraße 7/B2, 91058 Erlangen, Germany}

\begin{abstract} 
In loop quantum gravity, the area element of embedded spatial surfaces is given by a well-defined operator. We further characterize the quantized geometry of such surfaces by proposing definitions for operators quantizing scalar curvature and mean curvature. By investigating their properties, we shed light on the nature of the geometry of surfaces in loop quantum gravity. 

We also investigate the entanglement entropy across surfaces in the case where spin network edges are running within the surface. We observe that, on a certain class of states, the entropy gradient across a surface is proportional to the mean curvature. In particular, the entanglement entropy is constant for small deformations of a minimal surface in this case. 
\end{abstract} 
\maketitle


\tableofcontents

\section{Introduction} 
\label{se_intro}
The quantization of spatial geometry is a cornerstone of loop quantum gravity (LQG)
\cite{Thiemann:2007zz,Ashtekar:2004eh}. 
Operators for the volume of spatial regions and the area of spatial surfaces had been defined early on \cite{Rovelli:1994ge,Ashtekar:1996eg,Ashtekar:1997fb} and became important for the further development of the field. The quantization of the Hamiltonian constraint \cite{Thiemann:1996aw,Thiemann:1996av} and the quantum theory of isolated horizons \cite{Smolin:1995vq,Rovelli:1996dv,Ashtekar:1997yu,Engle:2010kt} are examples of this. 

The eigenstates for the spatial geometry are the spin network states. The picture that emerges is, broadly speaking, that the vertices of the spin networks of valence four and higher contribute volume and that the edges can be considered as flux-tubes of area. 
But the finer details of this picture remain unclear. This is partially because the theory of spatial geometry in LQG is a genuine quantum theory. For example, area operators for intersecting surfaces do not commute, and thus they cannot be simultaneously diagonalized. Hence, any classical picture of the geometry will have to be lacking in important aspects. A notable refinement of the picture came in the form of the proposal that vertices correspond to polyhedra in flat space \cite{Bianchi:2010gc}. Indeed, there is a close correspondence between SU(2) intertwiner spaces and the quantization of a certain phase space corresponding to such flat polyhedra. However, although the quantum theory based on the polyhedra picture is closely related to that in LQG \cite{Bianchi:2011ub}, they are not the same. How they are related in detail is currently an open question.
 
Another important aspect of the quantum theory of spatial geometry is the following: Since LQG is a quantum theory, the standard notions of quantum information theory apply (see for example \cite{Girelli:2005ii}). In particular, there is entanglement across surfaces, and one can define the corresponding entanglement entropy \cite{Donnelly:2008vx}. It was conjectured already in \cite{Donnelly:2008vx} that there should be a correspondence between entanglement entropy and geometric quantities. However, it remains unclear which geometric aspect of a surface or the bulk spaces that it divides is related to the entropy. 

In the present work, we want to investigate the kinematic quantum geometry of surfaces. 
We consider scalar curvature
\begin{equation}
\label{eq_R}
R=\twoD R_{abcd} \,\twoD g^{bd}\,\twoD g^{ac}
\end{equation}
and mean curvature 
\begin{equation}
\label{eq_K}
H=\frac{1}{2}\twoD K_{ab}\,\twoD g^{ab},  
\end{equation}
and their counterparts in the quantum theory. Here, $\twoD R$ and $\twoD g$ denote Riemann curvature tensor and metric of a two dimensional surface, and $\twoD K$ is the extrinsic curvature of its embedding into space.\\

Our motivation is threefold:
\begin{enumerate}
\item The geometric quantities \eqref{eq_R} and \eqref{eq_K} are interesting in their own right. For example, they can shed light on the question whether surfaces have symmetries, which is relevant to the calculation of black hole entropy. 

\item Whether the curvature of surfaces is a meaningful concept in the quantum theory can inform the broad picture of geometry in LQG. One also gains intuition for the quantization of more complicated geometric quantities, and guidance for the choice of semiclassical states. 

\item For surfaces that divide space into disconnected parts, one can define the corresponding entanglement entropy. This begs the question if, and if so, how, this entropy is related to the geometry of the surface. 
\end{enumerate}

In order to define operators for various curvature invariants, we will express them in terms of simpler geometric quantities for which operators are available in LQG.
Similar ideas have been used before to propose operators for the scalar curvature in three dimensions \citep{Alesci:2014aza,Nemoul:2018vtt}.\\

Some cautious remarks are in order: Often, the spatial geometry is not an observable in the sense of Dirac. Therefore, the properties of kinematical states and operators acting on them 
are not necessarily indicative of the situation after the constraints have been taken into account. This point has been discussed in a controversial fashion \cite{Dittrich:2007th,Rovelli:2007ep}. At least in loop quantum cosmology, the kinematical quantum geometry also describes the dynamical sector \cite{Kaminski:2007ew}. In addition, there are situations in which the geometry of a surface is a Dirac observable, for example, the area of an isolated horizon. Finally, one can deparametrize the theory classically using matter fields as reference systems \cite{Brown:1994py}. In LQG, this can lead to a theory in which the kinematical Hilbert space, as well as the geometric operators thereon, can become physical \cite{Giesel:2007wn,Giesel:2012rb}. In view of all this, it is fair to say, however, that a general statement about the physical validity of our results cannot be made.  
We will consider gauge invariant states, and the surfaces we consider could conceivably be defined by matter fields. We thus see no obstacle to extending our results to the space of diffeomorphism invariant states. But the Hamilton constraint is not taken into account. Alternatively, one can deparametize the theory \cite{Giesel:2007wn}, whereupon the results would become physical. 

\section{Background} 
In the following, we will briefly review some aspects of the geometry of surfaces, as well as the relevant properties of geometric operators in LQG. 
\subsection{Intrinsic geometry via small circles}

\subsubsection{Smooth geometries}
\label{ssec:Circ_and_curv}
In \citep{gray1974}, the formulas for the volume and the surface area of a small, n-dimensional geodesic ball have been given as a power series in its radius $\epsilon$. Specializing to $n=2$, we obtain
\begin{equation}
U(S_{\epsilon}) = 2 \pi \epsilon \left ( 1 - \frac{R}{12} \epsilon^{2} + \mathcal{O}(\epsilon^{4}) \right)
\label{eqn:CircOfGeoBallOnManifold}
\end{equation}
for the circumference and 
\begin{equation}
A(S_{\epsilon}) = \pi \epsilon^{2} \left( 1 - \frac{R}{24} \epsilon^{2} + \mathcal{O}(\epsilon^{4}) \right)
\label{eqn:AreaOfGeoBallOnManifold}
\end{equation}
for the area of a small disc in an arbitrary manifold. Here, $R$ denotes the Ricci scalar (or scalar curvature) at the center of the circle. Note that, in the flat (zero curvature) case, these formulas reduce to the standard expressions for circumference and area of a circle in Euclidean geometry. Consider now the combination
\begin{align}
4 \pi A(S_{\epsilon}) - U(S_{\epsilon})^{2} &= 4 \pi^{2} \epsilon^{4} R \left( \frac{1}{6} - \frac{1}{24} + \mathcal{O}(\epsilon^{2}) \right)\\ 
&= \frac{\pi^{2}}{2} \epsilon^{4} R + \mathcal{O}(\epsilon^{6}) \, .
\end{align}
Solving this expression for the Ricci scalar, we obtain
\begin{align}
R &= \frac{ 8 \pi A(S_{\epsilon}) - 2\, U(S_{\epsilon})^{2} + \mathcal{O}(\epsilon^{6})}{\pi^{2} \epsilon^{4}}\\
&= \frac{ 8 \pi A(S_{\epsilon}) - 2\, U(S_{\epsilon})^{2}}{\pi^{2} \epsilon^{4}} + \mathcal{O}(\epsilon^{2}) \, .
\end{align}
Therefore, in the limit of small radii, we are left with
\begin{equation}
\label{eq:Rclass}
R = \lim_{\epsilon \rightarrow 0} \frac{ 8 \pi A(S_{\epsilon}) - 2\, U(S_{\epsilon})^{2}}{\pi^{2} \epsilon^{4}} \, .
\end{equation}
At this point, there is an ambiguity in whether to express the fourth power of $\epsilon$ in the denominator in terms of the area or the circumference. In full generality, we can even write
\begin{equation}
\pi^{2} \, \epsilon^{4} = \frac{1}{\left( 4 \pi \right)^{2-2\alpha}} A^{2\alpha} U^{4 - 4\alpha} + \mathcal{O}(\epsilon^{5}) \, ,
\end{equation}
which holds for arbitrary (real) values of the parameter $\alpha$. Since we evaluate the right hand side of equation \eqref{eq:Rclass} in the limit of vanishing $\epsilon$, we are going to neglect the higher order corrections in the previous expression for $\pi^{2} \epsilon^{4}$. Substituting this back into equation \eqref{eq:Rclass} then leads to
\begin{equation}
R = \lim_{\epsilon \rightarrow 0} \left(4\pi\right)^{2-2\alpha} \, \frac{ 8 \pi A(S_{\epsilon}) - 2\, U(S_{\epsilon})^{2}}{A^{2\alpha} U^{4 - 4\alpha}}
\end{equation}
for the scalar curvature. Using this formula to calculate the scalar curvature of a sphere, i.e., inserting 
\begin{equation}
\label{eq:Usphere}
U = 2 \pi r \sin{\theta}
\end{equation}
and
\begin{equation}
\label{eq:Asphere}
A = 2 \pi r^{2} \left( 1 - \cos{\theta} \right)
\end{equation}
into the previous equation (where $\theta = \frac{\epsilon}{r}$, with $r$ the radius of the sphere and $\epsilon$ the radius of the circle around the point at which we want to determine the curvature), we obtain
\begin{equation}
R = \frac{8}{4^{\alpha} r^{2} \left( 1 + \cos{\theta} \right)^{2-2\alpha}} \, .
\label{eqn:ScalarCurvatureOfSphereWithAlpha}
\end{equation}
If we assume $\theta$ to be small (i.e., $\epsilon << r$), this formula approximates the curvature of a sphere of radius $r$ correctly for arbitrary choices of the parameter $\alpha$, as can be seen from the Taylor expansion
\begin{equation}
R \approx \frac{2}{r^{2}} \left[ 1 + \left( 1 - \alpha \right) \frac{\theta^{2}}{2} + \mathcal{O}(\theta^{4}) \right] \, .
\end{equation}
However, if we choose $\alpha = 1$, the dependence on $\theta$ drops out completely (which can also easily be seen from equation \eqref{eqn:ScalarCurvatureOfSphereWithAlpha}) and the formula then gives the exact expression for the curvature of a sphere irrespective of the radius of the circle we use to measure it. In the following, we will therefore set $\alpha = 1$ and work with
\begin{equation}
R = \lim_{\epsilon \rightarrow 0} \frac{ 8 \pi A(S_{\epsilon}) - 2\, U(S_{\epsilon})^{2}}{A(S_{\epsilon})^{2}} \, .
\label{eqn:ScalarCurvViaAreaAndCirc}
\end{equation}
Note that there is also another possibility to extract an expression for the scalar curvature in terms of $A$ and $U$ from equations \eqref{eqn:CircOfGeoBallOnManifold} and \eqref{eqn:AreaOfGeoBallOnManifold}. We can simply negelect the terms of order $\epsilon^{4}$ (and higher) in both equations and solve the truncated equations for the scalar curvature $R$ by eliminating the radius $\epsilon$. This approach leads to
\begin{equation}
R = \frac{6 \pi}{A} \left[ 1 - \frac{U^{2}}{8 \pi A} - \frac{U^{4}}{128 \pi^{2} A^{2}} - \frac{U^{3}}{16 \pi A^{2}} \, \sqrt{\frac{U^{2}}{64 \pi^{2}} + \frac{A}{2 \pi}} \right] .
\label{eqn:ScalarCurvComplicatedExpr}
\end{equation}
Just as equation \eqref{eqn:ScalarCurvViaAreaAndCirc}, this formula correctly yields vanishing curvature when inserting the expressions for area and circumference of a circle in flat geometry. Also, when inserting the relations on a sphere, the correct result of $R = \sfrac{2}{r^{2}}$ is obtained if the radius of the circle tends to zero. However, equation \eqref{eqn:ScalarCurvViaAreaAndCirc} has the benefit of exactly reproducing the curvature for the spherical case without the need of taking a limit. This advantage, together with the fact that it is also a considerably simpler expression, convinced us to use equation \eqref{eqn:ScalarCurvViaAreaAndCirc} rather than \eqref{eqn:ScalarCurvComplicatedExpr} for the purpose of constructing a corresponding quantum operator.\\ 

Classically, our formula for the scalar curvature \eqref{eqn:ScalarCurvViaAreaAndCirc} always yields an exact result. In the quantum theory, however, where geometry becomes discrete, we cannot guarantee the existence of a sufficiently small circle. Therefore, the quality of our approximation strongly relies on whether $\epsilon^2R$ (together with similar terms containing other curvature scalars, which appear at higher orders in \eqref{eqn:CircOfGeoBallOnManifold} and \eqref{eqn:AreaOfGeoBallOnManifold}) are small compared to unity. This means that the radius of curvature must be large compared to the radius of the circle. 
In order to evaluate this condition we introduce a quantity which we call the coverage and which is motivated from the positive curvature case as follows: At points with positive curvature, the radius of curvature describes the radius of the sphere that gives the best approximation of the surface under consideration at the given point. The condition for our approximation means that the circles we consider can only cover a small portion of that sphere. If we center the circles at the north pole of the sphere, the inclination angle $\alpha$ of any point on the circle gives a measure for how much of the sphere is covered by the circle.

Similar considerations can be applied to the hyperbolic case ($R < 0$), where we approximate the local geometry by that of a hyperboloid embedded in Minkowski space. 
With $r$ being the radius of curvature of the hyperboloid in its standard parametrization and $\epsilon$ being the intrinsic radius of a circle around the hyperboloid's pole, we define $\alpha=-\frac{\epsilon}{r}$. The net result is then the occurrence of the hyperbolic cosine instead of the trigonometric cosine in the formula for $\alpha$, which leads to the coverage no longer having the interpretation of an angle. 
Altogether, we obtain 
\begin{equation}
\AUA=
\begin{cases} 
\quad\arccos\left(\frac{U^2}{2\pi A}-1\right)& \text{ for } R>0\\
- \arcosh\left(\frac{U^2}{2\pi A}-1\right)& \text{ for } R\leq0
\end{cases} \, .
\label{eq:AUA}
\end{equation}
The minus sign in the case $R\leq0$ is an arbitrary choice that will later help to distinguish more clearly between surfaces that are positively or negatively curved, respectively. For the purposes for which we use $\alpha$, this sign has no direct mathematical relevance and hence, if we speak of $\alpha$ getting large or small in the following, we always refer to the modulus of $\alpha$, unless stated otherwise. In summary, the approximation we use to define the curvature is better, the smaller the value of $\alpha$ is.\\
Instead of starting from the expression for the area of a circle on a sphere/hyperboloid, we could have equally used the formula for its circumference (for the sphere see \eqref{eq:Usphere}) as a starting point. Pursuing this strategy leads to the same overall results. However, the details of the calculation differ and are a little more involved because of domain issues and the occurrence of a square root. We therefore decided to work with the expression \eqref{eq:AUA} obtained from the approach using the area formula.\\

Note that this is not the only significance of $\alpha$. Recall that there is a close relationship between curvature and topology for surfaces. For a closed surface $S$, we have the Gauss-Bonnet theorem 
\begin{equation}
\int_S R\, \text{d}A = 4\pi\chi ( S )
\label{eq:GaussBonnet}
\end{equation}
with $\chi ( S )$ being the Euler characteristic of $S$. For later use, we note that the contribution of a small geodesic disc $S_\epsilon$ to this integral is, according to \eqref{eqn:ScalarCurvViaAreaAndCirc}, given by 
\begin{equation}
\label{eq:chiC}
8 \pi - 2\, \frac{U(S_{\epsilon})^{2}}{A(S_{\epsilon})}.
\end{equation}
This contribution can also be expressed in terms of the coverage $\alpha$. For the case $R>0$, e.g., it is given by 
\begin{equation}
AR=4\pi(1-\cos\alpha) \, .
\end{equation}

\subsubsection{Cone-like geometries}
So far, we have considered surfaces that are globally smooth. As it is unclear if this is an appropriate description for quantum surfaces, we will now briefly discuss surfaces with a cone-like singularity. Lacking a suitable definition of general conical geometries, we will investigate circles on  cones with a flat metric away from the tip. We will assume that the formulas we obtain will - in the limit of small circles - also hold true in case of the metric not being flat. Since, also in the case of arbitrary smooth surfaces, circumference and area of a circle take approximately the same form as on maximally symmetric surfaces if they are sufficiently small, this assumption seems to be justified.\par 
A cone can be parametrized in terms of its defect angle $\lambda\in [0,\, 2\pi)$. We can also extend this definition to arbitrary negative $\lambda$, in case of which the underlying geometry will turn hyperbolic. In terms of circumference and area of a circle around the apex, the defect angle is in both cases given as
\begin{equation}
\lambda=2\pi - \frac{U^2}{2A} \, .
\label{eq:lambdaofUandA}
\end{equation}
In contrast to the surfaces we have studied so far, the cones have distributional curvature, i.e., they are flat away from their apex, while at the apex itself the curvature diverges. As a consequence, the integral on the left hand side of the theorem of Gauss-Bonnet \eqref{eq:GaussBonnet} is not well-defined a priori. However, we can circumvent this problem by replacing the conical surface within a small circle of radius $\epsilon_0$ around the singularity by a spherical/hyperbolical hat (for $\lambda$ positive/negative) while demanding a smooth transition between both manifolds\footnote{More precisely, we demand the circumference of a circle, as a function of its radius $\epsilon$, and its first derivative to equate on both sides of the cutoff $\epsilon=\epsilon_0$.}. One can show that this requirement fixes the surface area of the smooth hat to
\begin{equation}
A_{\text{hat}} = \pm r^2 \lambda \, ,
\end{equation}
where $r$ is the curvature parameter of the spherical/hyperbolical hat, and therefore it does not depend explicitly on the position $\epsilon_0$ of the cutoff. As we replaced the singularity by a smooth surface of constant curvature $\pm 2 r^{-2}$, the Gauss-Bonnet surface integral is now well-defined and only collects a contribution from the smooth hat. Hence, the Euler characteristic of the cone-like surface is - neglecting the boundary term - purely determined by $\lambda$.

\subsection{Mean curvature and area change}\label{ssec:meancurvclassical}
It is well known that one can obtain the mean curvature $H$, which is proportional to the trace of the extrinsic curvature of an embedded surface, by determining variations in the surface area due to displacements along a geodesic field \cite{Wald:1984rg}. Let us sketch the setup for the convenience of the reader. 

We consider a two-surface $S$ embedded in a spatial three-manifold $\Sigma$, the latter of which being equipped with a metric $g_{\alpha \beta}$. Use Gaussian normal coordinates in a neighborhood of $S$, i.e., the set $\lbrace e^1(p), \, e^2(p), \, \ell \rbrace$, where $e^a$ are coordinates on $S$ and $\ell$ is an affine parameter of geodesics orthogonal to $S$. $\ell$ can be used to define a family of surfaces $S_\ell$ by displacing $S$ along the geodesic field. Furthermore, let $\xi^\alpha(p)$ denote the tangent of the geodesic through $p$.  Then the reduced metric is $h_{\alpha \beta}=g_{\alpha \beta}-\xi_\alpha \xi_\beta$, and the intrinsic metric on $S$ is $\twoD g_{ab}=h_{\alpha \beta}e^\alpha_a e^\beta_b$, where the $e$ are the tangent vectors given by the coordinates on $S$. 
By definition, the extrinsic curvature $K_{\alpha \beta}$ measures the normal component of the covariant derivative of tangent vectors of a surface.  In the adapted coordinates we use, it can also be expressed as a partial derivative of $h_{\alpha \beta}$ along the geodesic field, i.e., $K_{\alpha\beta}=\frac{1}{2}\partial_\ell h_{\alpha \beta}$, and the mean curvature then equals half of its trace: $H=\frac{1}{2}h^{\alpha\beta}K_{\alpha\beta}$. We can relate $H$ to the change of the area element $\sqrt{\text{det}\twoD g_{ab}}$ as we displace $S$ along the geodesic field\begin{equation}
\begin{aligned}
\partial_\ell \sqrt{\text{det}\twoD g_{ab}} &= \sqrt{\text{det}\twoD g_{ab}} \, \twoD g^{ab} \frac{1}{2} \partial_\ell \twoD g_{ab} \\
&=  \sqrt{\text{det}\twoD g_{ab}} \, \twoD g^{ab} K_{\alpha \beta}e^\alpha_a e^\beta_b \\
&=  2 \sqrt{\text{det}\twoD g_{ab}} \, H \, .
\end{aligned}
\end{equation}
The calculation shows that the mean curvature is proportional to the relative change of the area element on the surfaces $S_\ell$:
\begin{equation}
H=\frac{1}{2  \sqrt{\text{det}\twoD g_{ab}}} \frac{\partial \sqrt{\text{det}\twoD g_{ab}}}{\partial \ell} \, .
\label{eq:extrinsicAreaDeri}
\end{equation}

\subsection{Area and length in loop quantum gravity}

As we have seen in section \ref{ssec:Circ_and_curv}, the Ricci curvature at a point can be calculated using an expression that depends only on the area and the circumference of a small circle around that point. If we want to use this formula to promote the curvature to an operator in the quantum theory, we will therefore need quantum operators representing area and length observables. Fortunately, there exist proposals for both of them in the loop quantum gravity literature.\\
An area operator for loop quantum gravity has already been introduced in the early years of the theory \citep{Rovelli:1994ge}, with the full details worked out in \citep{Ashtekar:1996eg}. Its action on a general cylindrical function $\Psi_\gamma$ is given by
\footnote{The numerical factors in \citep{Ashtekar:1996eg} differ from those used here, since the definition of $l_\text{P}$ is different.}
\begin{equation} 
A_S \Psi_\gamma = 4 \pi \beta \, l_\text{P}^2 \sum_\alpha \left[ \sum_{I_\alpha, J_\alpha} \kappa_{I_\alpha}\kappa_{J_\alpha} X^i_{I_\alpha}X^i_{J_\alpha} \right]^{\frac{1}{2}} \Psi_\gamma \, ,
\label{eq:areaOperator}
\end{equation}
where the first sum is over intersections $\alpha$ of the spin network graph $\gamma$ with the surface $S$ and the second sum runs over all pairs of edges $I_{\alpha}$, $J_{\alpha}$ incident at $\alpha$.  $l_\text{P}=\sqrt{\hbar G}$ denotes the Planck length, with $G$ being Newton's constant. 
The numerical factor $\kappa_{I}$ associated to the edge $I$ is $+1$ if the edge $I$ lies above the surface $S$, $-1$ if it is below $S$ and $0$ if it lies entirely within $S$ (or does not intersect $S$ at all, but since we only sum over edges incident at punctures, this case does not occur in the expression for the area operator). Lastly, assuming all the edges are outgoing at the vertices in the surface, the $X^{i}_{I_{\alpha}}$ denote the $i$-th component of the left-invariant vector field acting in the representation space associated to the edge $I_{\alpha}$. At a single puncture (possibly with multiple incident edges), the eigenvalues of the area operator are of the form
\begin{equation}
\begin{split}
\lambda_{\alpha} &=4\pi \beta \, l_\text{P}^2 \times \\ 
&\quad \sqrt{2j^u\left (j^u+1\right)+2j^d (j^d+1 )-j^{u+d}(j^{u+d}+1 )} \, ,
\end{split}
\label{eq:areaSpectrum}
\end{equation}
where $j^{u}$, $j^{d}$ and $j^{u+d}$ denote the spins obtained from coupling the spins of all incident edges lying above, below or within $S$, respectively. The eigenvalues of the area operator on the full surface are then given by the sum of the individual contributions of the punctures. Note that the coupled spins $j^{u}$, $j^{d}$ and $j^{u+d}$ depend on the intertwiner at the puncture.\\

A length operator for LQG was defined in \cite{Thiemann:1996at}. Acting on a function cylindrical on a graph $\gamma$, it takes the form 
\begin{equation}
L(s)=\frac{1}{8\pi l_\text{P}^2}\sum_{\nu \in V ( \gamma )} \sum_{\nu \in s_i} \sqrt{-8 \, \text{tr} ( [ h_{s_i},\, V_\nu ] [ h_{s_i}^{-1},\, V_\nu ] ) } \, .
\label{eq:defLengthOp}
\end{equation}
Here, the curve $s$, along which the length is taken, is broken up into pieces $s_i$ that intersect $\gamma$ only in one of their endpoints, and, if not already present, a vertex is added to $\gamma$ at the intersection point. $V_\nu$ denotes the volume operator acting at the vertex $\nu$. \cite{Thiemann:1996at} employs the Ashtekar-Lewandowski version \cite{Ashtekar:1997fb} of this operator. 

\begin{figure}[!htbp]
\begin{center}
\subfigure[\label{fig:circrad}Circumference vs. radius via the length operator]{\includegraphics[width=.9\columnwidth]{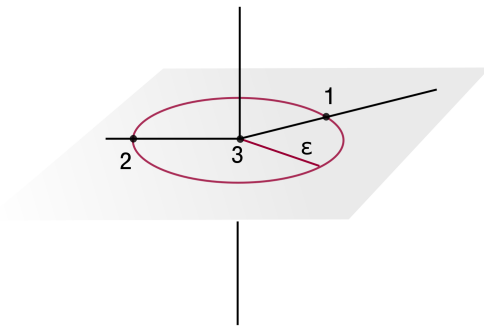}}
\subfigure[\label{fig:jujdjud}Circumference contribution as measured by Thiemann length operator]{\includegraphics[width=.9\columnwidth]{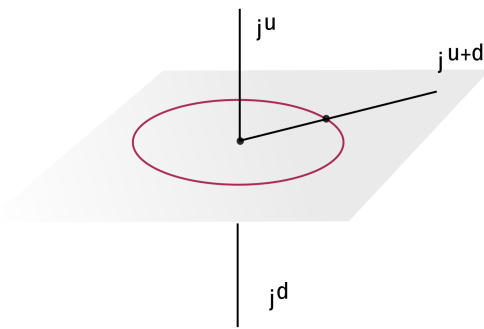}}
\end{center}
\caption{Vertex configurations}
\end{figure}

The length operator is important for the current work as it can be used to determine radius or circumference of small circles. Let us consider a circle around a vertex, as sketched in figure \ref{fig:circrad}. 
The radius $\epsilon$ receives a single contribution from the central vertex (marked 3 in the figure). The problem with the radius is that the action of the corresponding operator involves the action of the volume operator on a vertex of high valence (5-valent in the example in figure \ref{fig:circrad}), for which there is no known closed formula.

The circumference, on the other hand, receives potential contributions from the edges running out of the vertex, but inside the surface (marked 1,2 in figure \ref{fig:circrad}). However, these contributions vanish because of the properties of the AL volume operator, specifically, because of the fact that three linearly independent tangent vectors to edges are necessary to give a non-zero contribution. This can never be the case for vertices at which only edges running within the surface meet. A possible way out is to use the volume operator of Rovelli and Smolin  \cite{Rovelli:1994ge} (RS volume -- see \cite{Ashtekar:2004eh} for a definition in modern terms). 
However, there are two potential problems. One is that it is not clear whether the definition for the length operator \cite{Thiemann:1996at} goes through also with the RS volume, because of its different properties. The other problem is that the RS volume might be inconsistent with the semiclassical limit \cite{Giesel:2005bk,Giesel:2005bm}.

In the present work, we will make the choice to work with the length operator expressed in terms of the RS volume. We have checked that, for the circle in figure \ref{fig:jujdjud}, the length operator is well-defined and has the spectrum 
\begin{equation}
l=2\sqrt{8\pi\beta}\, l_P\sqrt[4]{\jud (\jud+1)} \, .
\label{eq:lengthSpectrum2valent}
\end{equation}
The more general case of a three-valent vertex, in which the length is evaluated along one of the edges, is also needed later. Again we have checked that the RS volume works in this case, and that the spectrum is unchanged from the original definition. It is given by  \cite{Thiemann:1996at}
\begin{widetext}
\begin{equation} 
\begin{aligned}
l = & \frac{\sqrt{8\pi\beta}\, l_P}{2\sqrt{j_3+\tfrac{1}{2}}}\sqrt{(j_3+1)\sqrt{(j_1+j_2+j_3+2)(j_1+j_2-j_3)(j_2+j_3-j_1+1)(j_3+j_1-j_2+1)}} \\
&\qquad \qquad \qquad \qquad \qquad \qquad \overline{ +j_3\sqrt{(j_1+j_2+j_3+1)(j_1+j_2-j_3+1)(j_2+j_3-j_1)(j_3+j_2-j_1)} } \, ,
\label{eq:lengthSpectrum3valent}
\end{aligned}
\end{equation}
\end{widetext}
where the representation $j_3$ is on the edge along which the length is measured. 

\subsection{Entanglement entropy in loop quantum gravity}
In the previous sections, we have focused on the geometric properties of surfaces in their own right. We have seen that their extrinsic geometry comprises information that cannot be obtained purely by measurements within the surfaces themselves. Instead, quantities requiring knowledge about the metric of the ambient space - at least in a vicinity of these surfaces - are inevitable. However, focusing on their embedding in space, their geometry is not the only aspect worth investigating. Consider a surface $S$ that separates the higher-dimensional bulk space into two disjoint regions. Then, according to information theory, these regions are possibly subjected to quantum entanglement and, in fact, this entanglement can be measured in LQG  \cite{Donnelly:2008vx}. Let $\Sigma$ be a spatial slice that is in a particular (pure) spin network state $\ket{\Psi_\gamma}$. The slice is divided into two regions $\Omega$ and $\bar{\Omega}$ by an arbitrary surface~$S$ which has no internal degrees of freedom (i.e., there are only trivial intertwiners sitting on the boundary). The spin network state can then be decomposed as a direct product of extended spin network states, defined on the subspaces $\Omega$ and $\bar{\Omega}$:
\begin{equation}
\ket{\Psi_\gamma}=\( \prod_{p=1}^P \frac{1}{\sqrt{2j_p+1}} \) \sum_{\vec{a}} \ket{\Psi_\Omega,\, \vec{a}} \otimes \ket{\Psi_{\bar{\Omega}},\, \vec{a}} \, .
\label{eq:SN_SchmidtDeco}
\end{equation}
The sum is understood to be running over all punctures $p$ at which $S$ is penetrated by edges of $\gamma$, and each vector $\vec{a}$ represents the expansion of the intertwiner at such a puncture in its $2j_p+1$ dimensional representation space. One can show that the states $\ket{\Psi_\Omega,\, \vec{a}}$ and $\ket{\Psi_{\bar{\Omega}},\, \vec{a}}$ are orthonormal and therefore \eqref{eq:SN_SchmidtDeco} can be interpreted as a Schmidt decomposition of the pure state $\ket{\Psi_\gamma}$. It is straightforward to calculate the entanglement in this case, and it turns out to be a logarithmic function of the edge spins penetrating $S$:
\begin{equation}
\ent(\Omega ) = \sum_{p=1}^P \ln{( 2j_p+1 )} \, .
\label{eq:SN_entropy}
\end{equation}
This entanglement entropy is symmetric, i.e., $\ent(\Omega)=\ent(\bar{\Omega)}$. Furthermore, as calculations are performed in the basis of spin network states that are also eigenstates of the known geometric operators, we are able to measure the entanglement between the subspaces as well as the geometric properties of the surface $S$ separating them at the same time. For our later purposes, we will extend the method presented here in section \ref{sec:entropymeancurvature}, allowing for $S$ to carry internal degrees of freedom.

\section{Quantum geometry of surfaces} 
\label{sec_quantgeo}

\subsection{Scalar curvature}
\subsubsection{Scalar curvature operator}

\begin{figure}[!htbp]
\begin{center}
\includegraphics[width=.9\columnwidth]{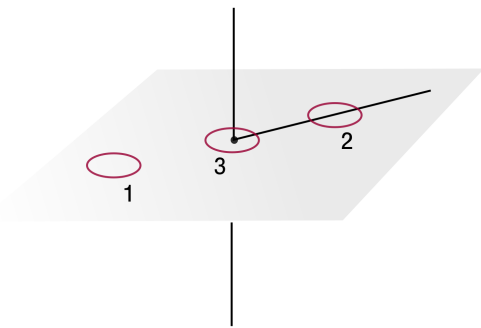}
\end{center}
\caption{Three possible configurations for circles on a surface in a given spin network state}
\label{fig:conf}
\end{figure}

Now, after all the classical considerations about intrinsic and extrinsic curvature of surfaces, and after having introduced the basic geometric operators of loop quantum gravity, we are finally ready to define quantum operators corresponding to these classical quantities. Let us start with the scalar curvature operator. A classical formula for this quantity was given in equation \eqref{eqn:ScalarCurvViaAreaAndCirc}. As already stated there, this formula expresses the scalar curvature at a point as a limit over shrinking circles centered at that point and it only depends on the area and the circumference of these small circles. Having introduced quantum operators corresponding to length and area in the previous section, we could simply replace these quantities in the classical expression by their corresponding quantum operators, leading to
\begin{equation}
\widehat{R}_{p} \coloneqq \lim_{r_{c_{p}} \rightarrow 0} \frac{8\pi}{\widehat{A}(c_{p})}-\frac{2\, \widehat{U}^2(c_{p})}{\widehat{A}^2(c_{p})} \, ,
\label{eq:KOperator}
\end{equation}
where $c_{p}$ represents a circle centered around $p$ and $r_{c_{p}}$ denotes its radius. Unfortunately, however, the fact that the kernel of the area operator is non-empty forces us to be a little more careful with the definition of this operator. For the purpose of this discussion, as well as throughout the rest of the paper, we will always work with eigenstates of both, the area operator and the length operator measuring the circumference of a circle around the point under consideration. This is possible since the area operator and the circumference operator act at different points of the spin network and therefore commute with each other. Given a specific circle in our manifold and an arbitrary eigenstate of the circle's area and circumference operators, we can have either of the following three situations:
\begin{enumerate}
\item	The eigenvalues of both area and circumference operator in this state are zero.
\item	The area eigenvalue equals zero but the circumference operator has non-zero eigenvalue.
\item	The area operator has non-vanishing eigenvalue.
\end{enumerate}
These are schematically depicted in figure \ref{fig:conf}.

The third case is the simplest, since it allows us to directly use equation \eqref{eq:KOperator} to obtain a finite value for the curvature. Note that, in this case, curvature is always concentrated at punctures since the definition of the curvature is given as a limit of shrinking circles (keeping their center fixed) and thus they only have non-vanishing area if there is a puncture in the center of the circle.\\
The second case describes a degenerate circle and occurs if the only spin network edges that touch the circle are lying entirely within the surface. Obviously, the expression for the curvature diverges in this case. One way to resolve this issue is to define a constant $c_1$ and use it to associate a finite curvature to points where this situation occurs. Another option would be to leave the scalar curvature operator undefined on states for which this situation occurs at the point under consideration. However, for a given point $p$, this would make it necessary to pay close attention to what states the operator $\widehat{R}_{p}$ is defined on and expressions containing the action of $R_{p}$ on a spin network state would never be well-defined without fixing the point $p$. We will therefore stick to the first option.\\
The first case is the most generic since most circles in the spatial manifold will not touch the spin network at all. In this case, the circumference and the area are both equal to zero. This again leads to a divergent expression for the curvature and we will deal with this divergence in the same way as for the second case, but using a different constant $c_{2}$.\\
Note that, in the first and second case, we are considering circles that have no area. Classically, such degenerate circles are not capable of detecting curvature. However, in classical geometry we can find arbitrarily small circles around every point that have non-vanishing area. This is not the case in the quantum geometry described by loop quantum gravity. Therefore, it seems natural to choose $c_1 = c_2 = 0$ and ascribe curvature only to such points, where a spin network edge punctures the surface. With this choice, the full scalar curvature operator is then given by
\begin{equation}
\widehat{R}_{p} \coloneqq 
\begin{cases}
\lim_{r_{c_{p}} \rightarrow 0} \frac{4\pi}{\widehat{A}(c_{p})}-\frac{\widehat{U}^2(c_{p})}{\widehat{A}^2(c_{p})} &\quad \text{if $p$ is a puncture} \\
0 &\quad \text{otherwise}
\end{cases} \, .
\end{equation}

Finally, we consider the quantization of the coverage angle  \eqref{eq:AUA}.  
Its quantum version can be obtained by substituting $U$ and $A$ with their respective quantum operators. 
The same divergences as in the case of the scalar curvature operator also show up here, and, in analogy to the scalar curvature operator, we choose the coverage angle operator to be non-zero only at punctures.
At a puncture, the general coverage operator is therefore defined as  
\begin{equation}
\widehat{\alpha}_{p} \coloneqq \lim_{r_{c_{p}} \rightarrow 0} \begin{cases} 
\quad\arccos\left(\frac{\widehat{U}(c_{p})^2}{2\pi \widehat{A}(c_{p})}-1\right)& \text{ for } R_{p}>0\\
-\arcosh\left(\frac{\widehat{U}(c_{p})^2}{2\pi \widehat{A}(c_{p})}-1\right)& \text{ for } R_{p}\leq0
\end{cases} \, .
\label{eq:alphoperator}
\end{equation}
In the section on curvature, we also mentioned the possibility of conical curvature singularities. In this case, curvature is quantified by the defect angle \eqref{eq:lambdaofUandA} which can be quantized analogously to the scalar curvature, i.e., by replacing area and circumference with their corresponding quantum operators and treating the kernel of the area operator in the same way as before. This leads to
\begin{equation}
\widehat{\lambda}_{p} \coloneqq \begin{cases} \lim_{r_{c_{p}} \rightarrow 0} \left( 2\pi - \frac{\widehat{U}(c_{p})^2}{2\widehat{A}(c_{p})} \right) & \quad \text{if p is a puncture}\\
0 & \quad \text{otherwise}
\end{cases}
\label{eq:lamboperator}
\end{equation}
as an expression for the defect angle operator in the quantum theory. 

Finally, we can also give an expression for the Gauss-Bonnet integral in the quantum theory:
\begin{equation}
\label{eq:quantumgb}
\int_S \widehat{R}\, \text{d}\widehat{A} =
\sum_p \lim_{r_{c_p}\rightarrow 0} \left[8 \pi - 2\, \frac{\widehat{U}(c_p)^{2}}{\widehat{A}(c_p)}\right]\, , 
\end{equation}
where the sum over $p$ denotes a sum over punctures.

\subsubsection{Spectrum and physical implications}

\begin{figure}[!htbp]
\begin{center}
\includegraphics[width=.9\columnwidth]{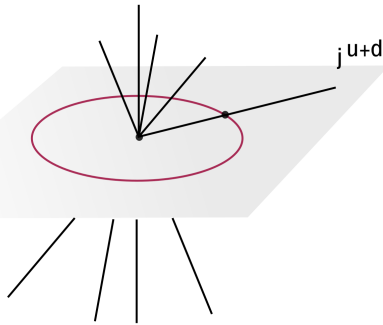}
\end{center}
\caption{\label{fig:vertex}Vertex type for the discussion of intrinsic curvature}
\end{figure}

In the following, we will discuss the spectrum of the intrinsic geometry for special spin network configurations. We will consider a puncture that has at most one holonomy running tangentially, but is otherwise generic, like the one depicted in figure \ref{fig:vertex}.  
We can then choose a recoupling scheme that makes the vertex effectively three valent, as in figure \ref{fig:jujdjud}. In the following, we will consider a basis element from this recoupling scheme. We will denote the spins as in figure \ref{fig:jujdjud}: $\ju,\,\jd$ and $\jud$. These spins have to fulfill the coupling rules for angular momenta. 

The first case we consider is $\jud=0$. Then $\ju=\jd$ and the curvature eigenvalues scale with $j^{-1}$. In this case, however, $\alpha$ and $\lambda$ take values of $\pi$ and $2\pi$, respectively, which corresponds to full coverage. 

\begin{figure}[!htbp]
\begin{center}
\includegraphics[width=\columnwidth]{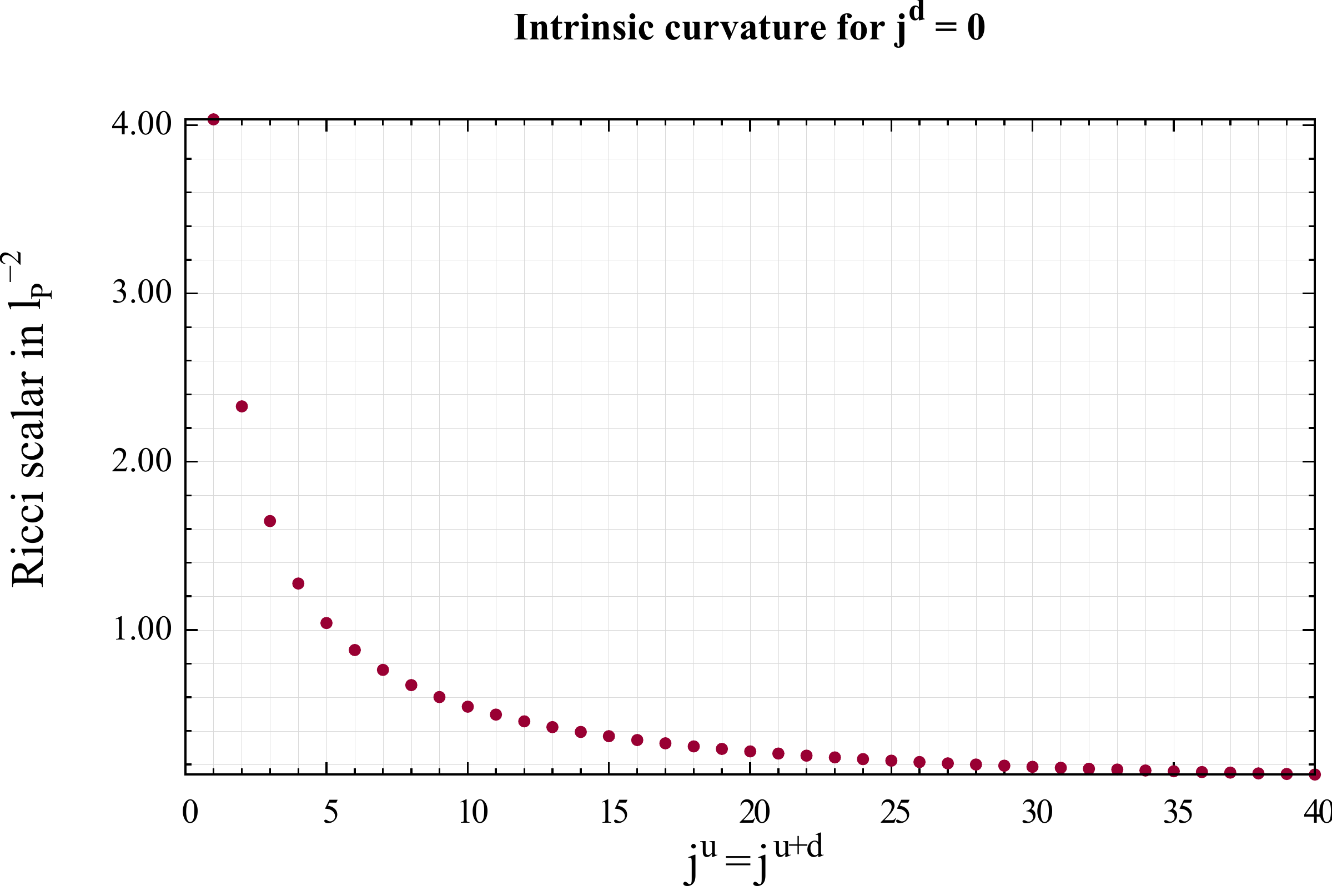}
\end{center}
\caption{This plot shows the Ricci scalar $R$ in units of $l_P^{-2}$ in the case where $\jd=0$. In this case, we necessarily have $\ju=\jud$, which allowed us to reduce the plot by one dimension.}
\label{fig:intrinsic40-0}
\end{figure}

Next, we discuss vertices with one transversal edge and no edge below the surface, i.e., $\jd = 0$ and $\ju = \jud$. These might be associated with black hole horizons, as the latter are often modeled as a boundary of space-time so that all spin networks in the interior have to vanish. Figure \ref{fig:intrinsic40-0} shows the behavior of the curvature operator applied to these two-valent vertices. The curvature eigenvalue tends to zero for large spins. 
On the other hand, it turns out that these vertices have a constant coverage and defect angle, independent of the corresponding edge spins. One obtains 
\begin{equation}
\alpha=\arccos\left( \sfrac{4}{\pi}-1 \right)\qquad \lambda=2\pi-4. 
\end{equation}

Finally, let us look at the case where all three spins are non-vanishing. For $\jud$ taking the maximal value allowed by coupling, while still having $\ju=\jd$, the circumference becomes large with growing spin, implying that the underlying geometry turns hyperbolic. While the curvature tends to a constant negative value, coverage and defect angle grow to negative infinity. Their absolute value is small only in a narrow regime of spins. 

\begin{figure}[!htbp]
\begin{center}
\includegraphics[width=\columnwidth]{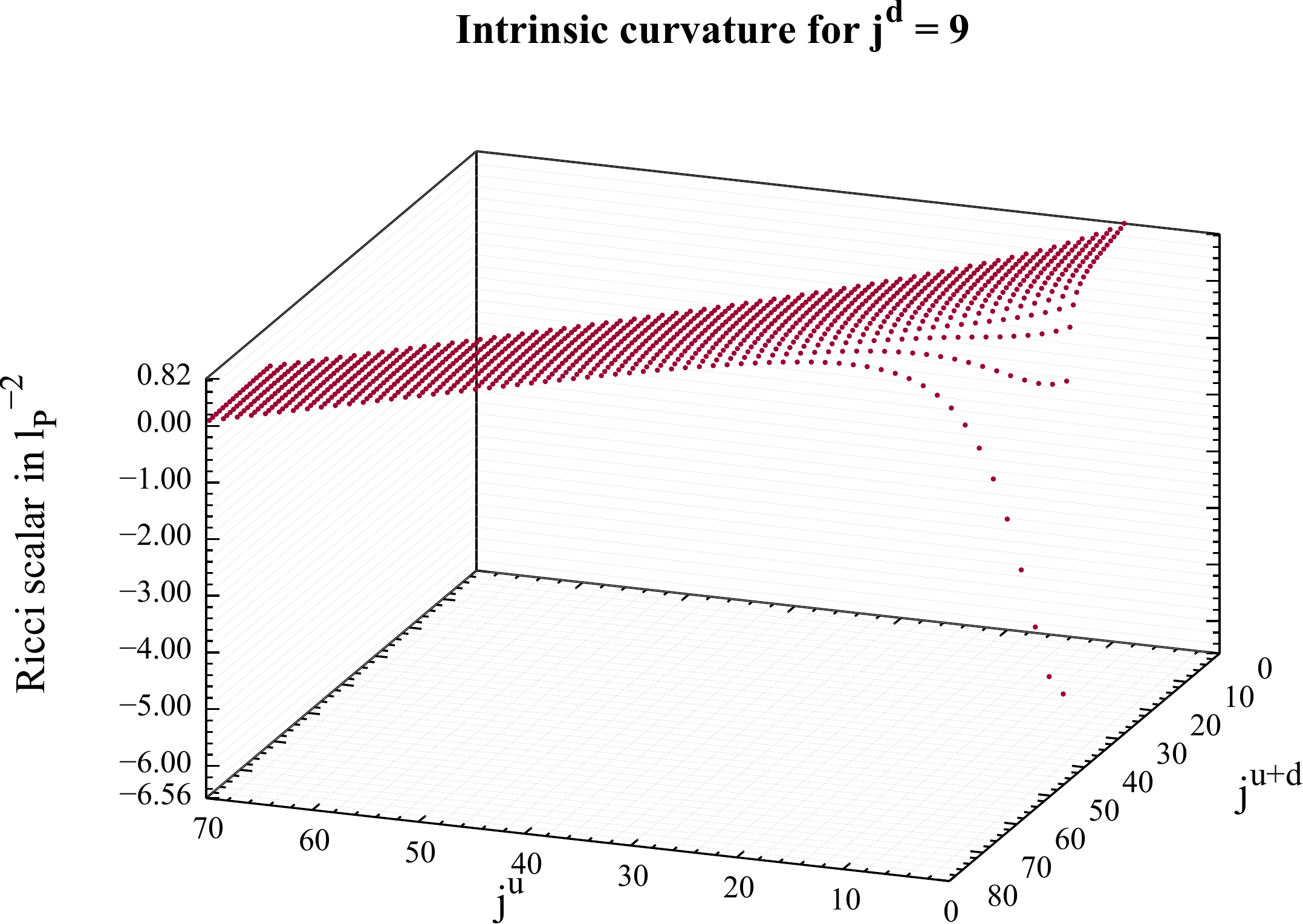}
\end{center}
\caption{The Ricci scalar $R$ was plotted for a fixed value of $\jd=9$ as a function of $\ju$ and $\jud$. It shows the behavior for $\ju$ growing large compared to $\jd$.}
\label{fig:intrinsic70-9}
\end{figure}

For the cases of low spins ($j\lesssim 10$), numerical calculations show that the curvature of a single puncture is quite large, i.e., of the order of several inverse Planck lengths squared. However, the surface becomes flatter if at least one of the spins $\ju,\jd$ grows. This is shown in figure \ref{fig:intrinsic70-9}. One can also observe that, as long as $\jud$ takes moderate values, the tangential spin does not have a big influence on the curvature. When $\jud$ grows large, however, the surfaces become strongly hyperbolic. 

\begin{figure}[!htbp]
\begin{center}
\includegraphics[width=\columnwidth]{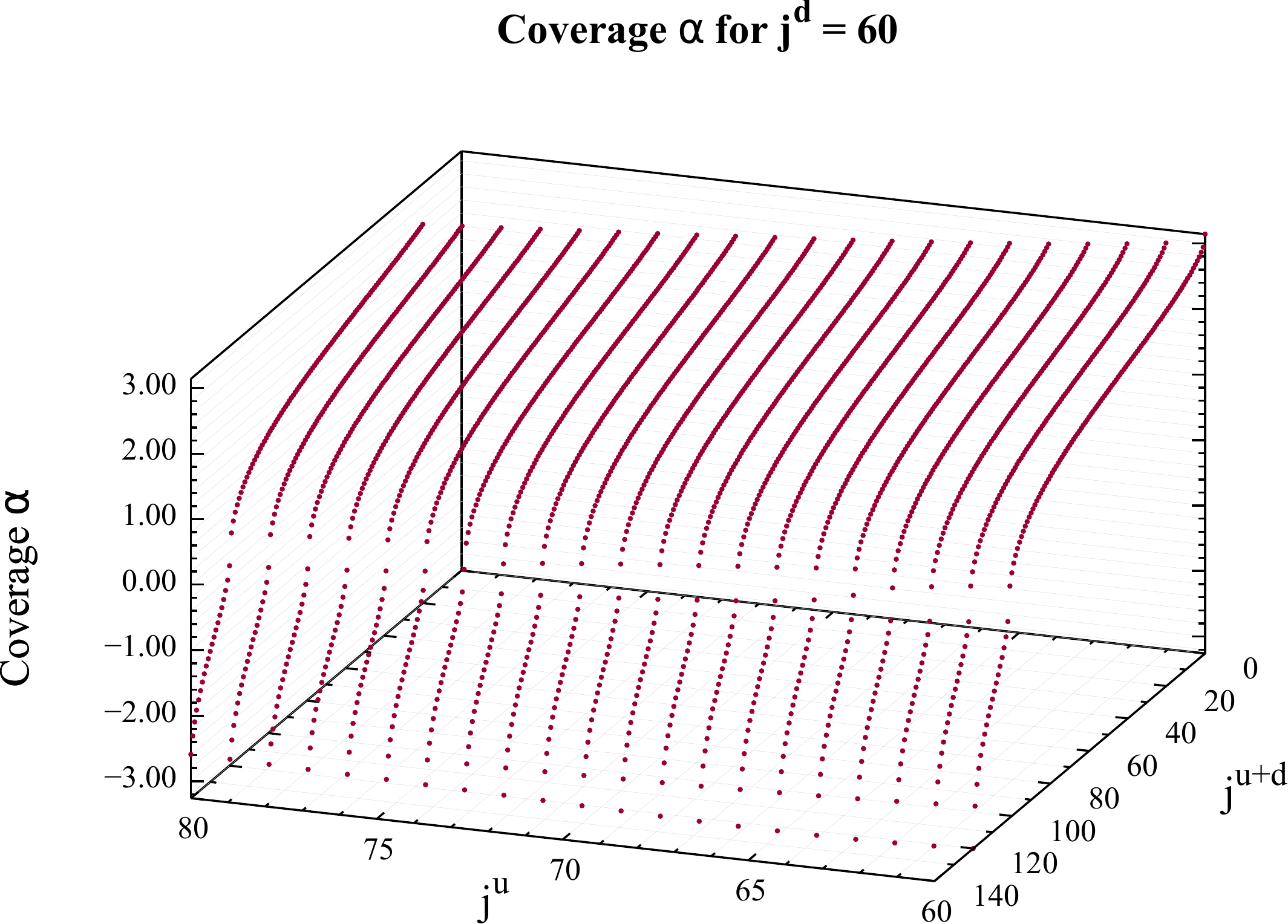}
\end{center}
\caption{The coverage angle $\alpha$ was plotted for a fixed value of $\jd=60$ as a function of $\ju$ and $\jud$. The range was chosen such that the maximal $\ju$ is still of the same order of magnitude as $\jd$.}
\label{fig:coverage80-60}
\end{figure}

\begin{figure}[!htbp]
\begin{center}
\includegraphics[width=\columnwidth]{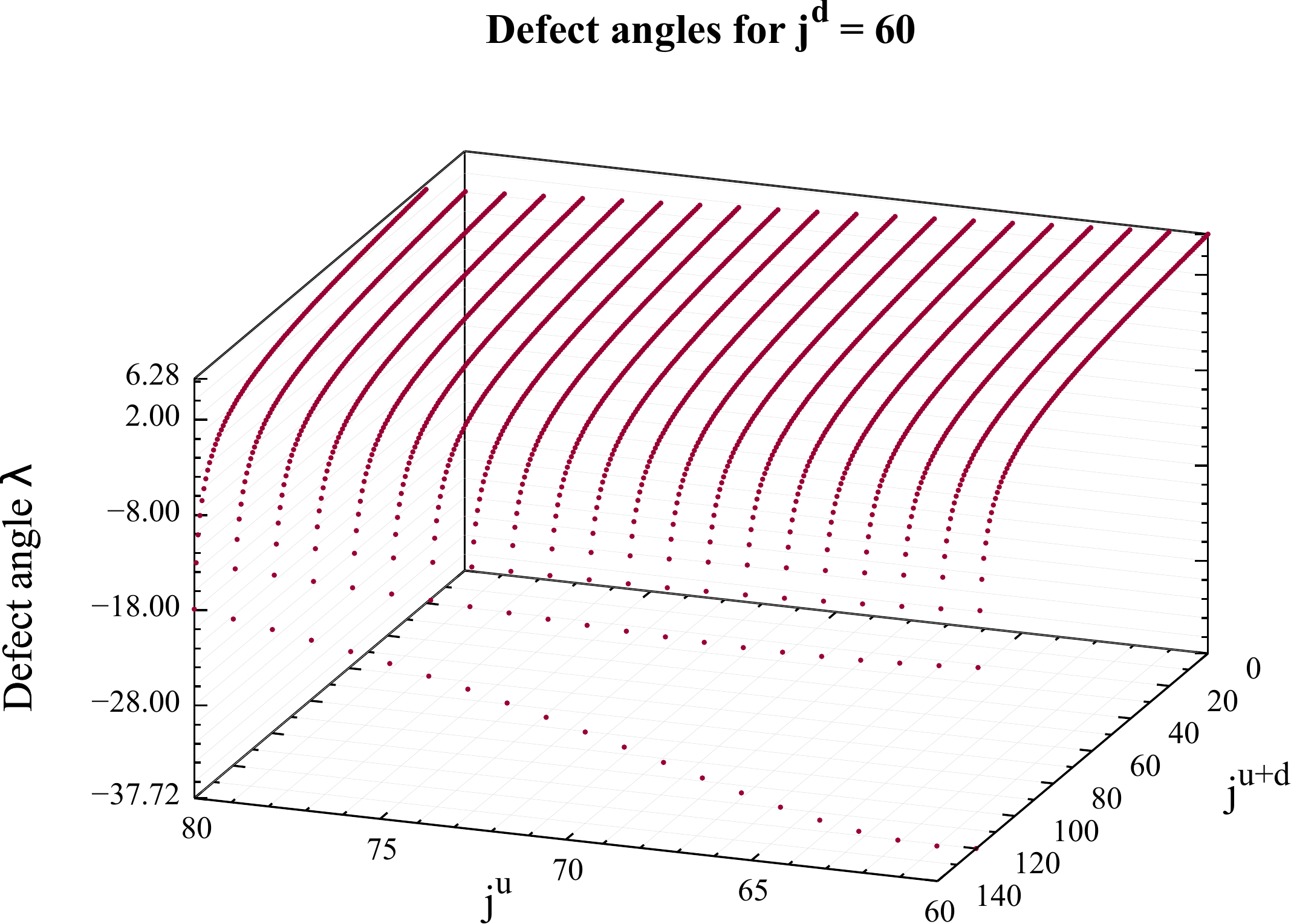}
\end{center}
\caption{Defect angle $\lambda$ plotted for a fixed value of $\jd=60$ as a function of $\ju$ and $\jud$. The range was chosen such that the maximal $\ju$ is still of the same order of magnitude as $\jd$.}
\label{fig:Cone80-60}
\end{figure}

Figures \ref{fig:coverage80-60} and \ref{fig:Cone80-60} show the coverage and defect angle in a slightly different range of quantum numbers. Both quantities behave qualitatively similar. They show a stronger dependence on the tangential spin than the curvature. 
In the case of the coverage, there is a gap around zero which means that, considering the smooth geometries, there are even fewer states for which the ratio $\sfrac{\epsilon}{r}$ is small. 

Our results suggest that single punctures on a two-surface possess an intrinsic curvature that is often large. The generically large curvatures are worrisome as it is unclear how the strongly curved patches can connect to form a macroscopic surface that is -- if at all -- curved only on much larger scales. There are regimes of low curvature, however, most prominently in the limit of high spins carried by the transversal edges. 

Nevertheless, we have to question the extent to which the approximations we used to obtain the presented geometric operators are valid. As discussed in section \ref{ssec:Circ_and_curv}, the classical formulas for the scalar curvature require the limit in which the radius $\epsilon$ of the circles becomes small compared to the radius of curvature $r$. This ratio is measured by the coverage parameter $\alpha=\pm\sfrac{\epsilon}{r}$. Unfortunately, most of the states of the three valent vertex yield large coverages that even exceed unity. Hence, $\widehat{R}_p$ can probably not be considered meaningful in those regimes. On the other hand, there are states (for transversal spins of same order and with intermediate $j^{u+d}$) for which the coverage $\alpha$ becomes small.

In the picture of a conical geometry, we have used the defect angle $\lambda$ to characterize the geometry of the vertex. $\lambda$ is similar to the coverage $\alpha$. Qualitatively, both quantities show a similar dependence on the quantum numbers. From our results, it is not clear whether the conical picture or the smooth picture is more appropriate to describe the quantized geometry. 

$\alpha$ and $\lambda$ also represent a direct measure of the contribution of 
a small disc to the left hand side of the Gauss-Bonnet theorem \eqref{eq:GaussBonnet}. We can investigate the quantized version \eqref{eq:quantumgb}, for example, by asking how many punctures are needed to form a closed topological sphere, i.e., to obtain $\chi(S)=2$ in the quantum theory. Assuming, for simplicity, that all punctures are in the same configuration, we obtain, depending on the concrete state, a required number of punctures in the order of $10^1$ with the result not necessarily being an integer. This is again not quite what one would expect from physical intuition. It also contradicts results on black holes \cite{Domagala:2004jt} which show that macroscopic black holes have high numbers of punctures. 

One can also examine the contributions of higher valent vertices to the Gauss-Bonnet integral.  Adding more \emph{transversal} edges does not change the picture, as only the recoupled spins $\ju$, $\jd$ are relevant. On the other hand, the operator for the circumference $U$ is sensitive to the number of \emph{tangential} edges within the surface. However, it turns out that, already for few tangential edges, the geometry becomes strongly hyperbolic. Hence, these states do not seem to have properties, that are physically more appealing. In summary, our results suggest that macroscopic surfaces are rather formed by few spin network edges carrying a large spin than the other way round. On the other hand, the regime with large coverages of single punctures is not a valid domain for the approximations we have made. Therefore, the quantum geometry in these domains remains to be studied in a different way.

\subsection{Mean curvature}
\begin{figure}[!htbp]
\begin{center}
\includegraphics[width=\columnwidth]{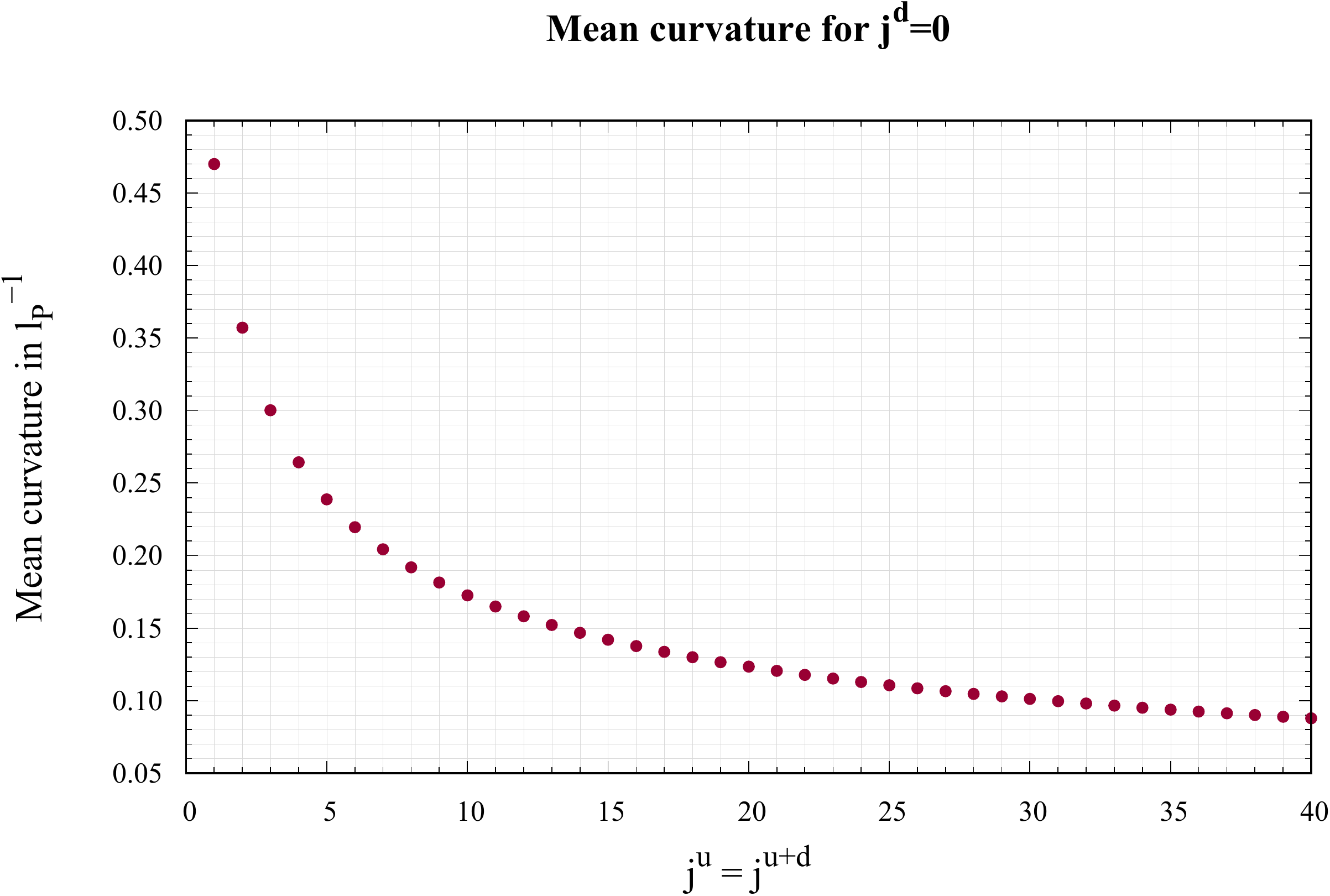}
\end{center}
\caption{This plot shows the scalar extrinsic curvature in units of $l_P^{-1}$ for the case where $\jd=0$. Here, we necessarily have $\ju=\jud$ which allowed us to reduce the plot by one dimension.}
\label{fig:extrinsic40-0}
\end{figure}

\begin{figure}[!htbp]
\begin{center}
\includegraphics[width=\columnwidth]{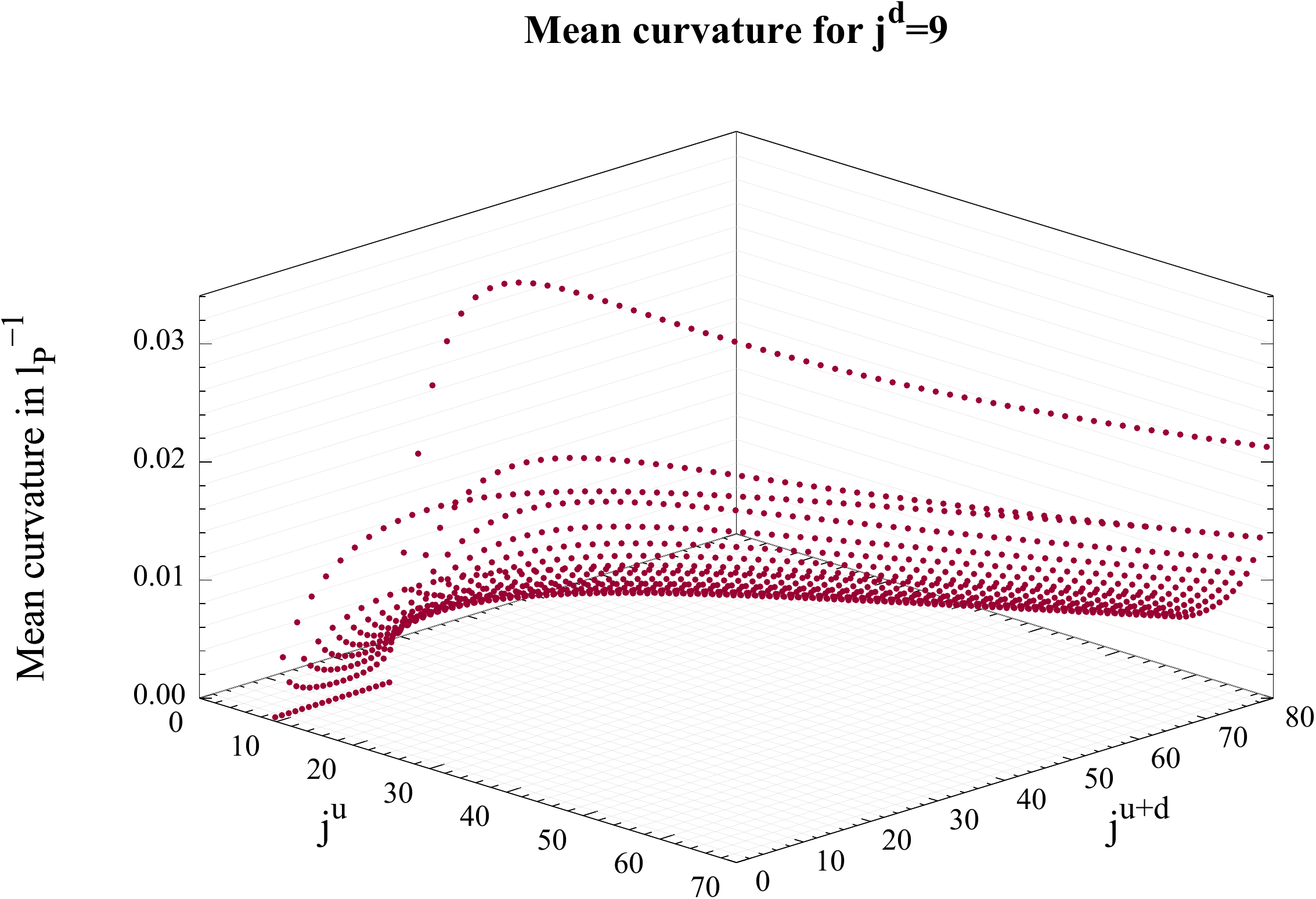}
\end{center}
\caption{This plot shows the scalar extrinsic curvature in units of $l_P^{-1}$ for fixed $\jd=9$.}
\label{fig:extrinsic70-9}
\end{figure}

In this section, we want to present an approach to quantize the mean curvature that we introduced in chapter \ref{ssec:meancurvclassical}. The main difficulty in this task will be to formulate the continuous partial derivatives of the classical theory in a fashion suitable to discrete quantum geometry. We will follow the obvious ansatz and express them in terms of difference quotients. The goal is to define an operator for the mean curvature at the point $p$ on a surface $S$, embedded in three-space $\Sigma$. To this end, we choose a family of surfaces $\{S_t\}_{t\in\mathbb{R}}$ that foliates a neighborhood of $S$ in $\Sigma$. We ask that $S_0=S$ and that the family is continuous in $t$ and each member $S_t$ of the family is homeomorphic to $S$ via a map
\begin{equation}
 \varphi_t: S \longrightarrow S_t \, . 
\end{equation}
Let $D(\epsilon)$ be a family of  topological discs in $S$ around $p$ that shrink to $p$ for $\epsilon\rightarrow 0$. We can also define the image
\begin{equation}
D(\epsilon,t)= \varphi_t(D(\epsilon))
\end{equation}
of the disc in the surfaces of the foliation. Finally, let $l(t)$ denote a suitably defined geometric length from $p$ to its image $\varphi_t(p)$. 
Then we can define the mean curvature operator as
\begin{equation}
\widehat{H}(p)\coloneqq \lim_{\epsilon \rightarrow 0} \, \lim_{t \text{\tiny{$\searrow$}} 0} \, \frac{1}{2\widehat{A} ( D(\epsilon) )} \frac{\widehat{A} (D(\epsilon,t )) - \widehat{A} ( D(\epsilon,-t ))}{\widehat{l} ( t )+\widehat{l} ( -t )} \, . \label{eq:defMeanCurvOp}
\end{equation}
Obviously, we again face the problem of the area operator in the denominator, which leads to $\widehat{H}(p)$ only being defined on spin network states having at least one transversely intersecting spin network edge through $p$. In the following, instead of carefully introducing constants to obtain a well-defined operator on the whole Hilbert space, we want to focus on states where $H$ yields meaningful results, i.e., we will only discuss its action on vertices with at least one edge transversely intersecting $S$.\\
Because of the properties of the operators involved, this definition is independent of the details of $\varphi_t$ and $D(\epsilon)$.
However, we still need to define the distance operators $\widehat{l}(\pm t)$ in a suitable fashion and examine their kernel and their commutation relations with the area operator. In \eqref{eq:shiftdistances}, we summarized a few possible definitions for $l(\pm t)$:
\begin{equation}
l(\pm t) \coloneqq
\begin{cases}
\min{\( l(e \in \mathcal{E}_{b/a}) \)} & \text{case (a)} \\
\frac{1}{N_{b/a}}\sum_{e\in \mathcal{E}_{b/a}} l(e) & \text{case (b)} \\
l\(\gamma_{b/a}(p)\) & \text{case (c)}
\end{cases} \, .
\label{eq:shiftdistances}
\end{equation}
Here, $\mathcal{E}_{b/a}$ denote the sets containing the transversal edges through $p$ below/above $S$ and $N_{b/a}$ their cardinalities. Furthermore, $\gamma_{b/a}$ is a transversal path through $p$ below/above $S$ which is not contained in and does not intersect any of the edges in $\mathcal{E}_{b/a}$. Now, having proposed multiple definitions of the shift distance, we need to figure out whether, and if so, how, they behave differently in the situations we consider. In case they do, we have to decide which definition we are going to work with.\\
In the first two of the above cases, the shift distance is measured along holonomies coinciding with the transversal edges through $p$. In the classical case, we have to displace the surface along a geodesic path. Thus, case (a) seems to provide a reasonable definition, as it highlights the properties of geodesics to extremize the length functional. However, for vertices with higher valence, we do not see any reason why two different length operators acting at the same vertex should commute with each other, and therefore case (a) does not yield a well-defined expression. Case (b) circumvents this problem since it does not require simultaneous diagonalization of multiple length operators. Therefore (b), which determines the average length measured along all transversal edges, is a well-defined expression. If both, $\mathcal{E}_{b}$ and $\mathcal{E}_{a}$, contain at most one element, then the definitions (a) and (b) agree. In case (c), the shift distance is measured along a path that does not coincide with any of the edges through $p$. In this case, $l(t)$ is necessarily symmetric, i.e., $l(t)=l(-t)$. Just like (b), (c) also yields a well-defined expression. However, we want to apply the mean curvature operator to the three-leg spin network configuration (cf.\ figure \ref{fig:jujdjud}) in the following. Here, measuring $l(\pm t)$ in the manner of (c) would yield an effectively four-valent problem for which we cannot determine the spectrum of the length operator explicitly. For the reasons summarized here, we will proceed using definition (b) to measure $l(\pm t)$ in the case of non-empty sets $\mathcal{E}_{b/a}$. On the other hand, if one of the $\mathcal{E}_{b/a}$ does not contain any edges, we choose to measure the shift distance on the corresponding side along an arbitrary path, as in case (c). Provided there is at least one edge intersecting the point $p$ transversely, $p$ is at least divalent as a vertex, and therefore this definition ensures that there is a non-vanishing contribution to the total shift distance in the low-valent case. Additionally, we know that, for valences less or equal to three, the spin network functions are eigenstates of the length operator, and therefore $A(c)$ and $l(\pm t)$ commute. Hence, in this case, the expression in \eqref{eq:defMeanCurvOp} remains well-defined and finite.\\
Instead of further dealing with the domains in Hilbert space that are critical to our operator, we want to move ahead at this point and apply it to states where we get meaningful results, in order to analyze its spectrum. In order to allow for comparison with the intrinsic properties of quantum surfaces, we will work in the $\ju,\, \jd,\, \jud$ representation of three-valent vertices we introduced before, and which is depicted in figure \ref{fig:jujdjud}. In contrast to the intrinsic operators, $H$ is sensitive to the orientation of the surface of interest in space (namely, $H$ is antisymmetric with respect to reversing it). In the following, we will, without loss of generality, restrict ourselves to the case where $\ju \geq \jd$, which corresponds to choosing the orientation of the surface such that its mean curvature is positive (or zero). A surface with vanishing mean curvature is also called a minimal surface. One trivial feature of our extrinsic curvature operator is to have vanishing eigenvalues on the $2\ju+1$ times degenerate states where the transversal spins coincide, i.e., $\ju=\jd$. Apart from these generic minimal surfaces, flat embedding in ambient space can also be achieved asymptotically for large quantum numbers. An important example is again the black-hole vertex with $\jd=0$. Here, we necessarily measure a constant ratio of two between the areas of the shifted and the actual surface, whereas the distance they are shifted by scales linearly with the quantum number $\ju=\jud$. The effectively hyperbolic decrease of $H$ with growing spins on these black hole horizons is shown in figure \ref{fig:extrinsic40-0}.
On the other hand, if the spin network below the surface does not vanish, the behavior of the mean curvature becomes more complicated. Figure \ref{fig:extrinsic70-9} shows a plot where we fixed the spin of the edge below the surface to $\jd=9$. As expected, we observe that, at first, $H$ grows rapidly as the difference between the transversal spins starts to increase. Also, the extrinsic curvature grows large when $\jud$, i.e., the spin of the tangential edge, takes values near its minimum and maximum (as allowed by recoupling theory), whereas it takes its smallest values for intermediate tangential spins. If we look at the regime where $\ju$ grows even larger when compared to $\jd$, the shifted area below the surface becomes more and more negligible and we asymptotically recover the black hole case. Hence, as already for the intrinsic case, we can state that it is sufficient to have one of the transversal spins grow large in order to have mean curvature tend to zero.

\section{Entropy and mean curvature} \label{sec:entropymeancurvature}
In our picture of surfaces in the quantum theory, the surfaces themselves can carry two kinds of geometrical excitations: the punctures on the one hand, and spin network edges running within the surfaces on the other. Because of the latter, the picture of entanglement entropy has to be slightly refined from the one in \cite{Donnelly:2008vx}. 

We consider a surface $S$ dividing $\Sigma$ into two disjoint parts $\Omega$ and $\bar{\Omega}$, i.e.,
\begin{equation}
\Sigma=\Omega\, \dot{\cup}\, S \, \dot{\cup}\,\bar{\Omega} \, , \qquad \partial \Omega=\partial\bar{\Omega} =S. 
\end{equation}
Thus, we have a tripartite division of the system where each of $S$, $\Omega$, $\bar{\Omega}$ is entangled with the other two. The corresponding entanglement entropies will be denoted $\ent_{S}$, $\ent_\Omega$, and $\ent_{\bar{\Omega}}$, respectively. In particular, 
\begin{equation}
\label{eq_nonunique}
\ent_\Omega\neq \ent_{\bar{\Omega}}
\end{equation}
in general, i.e., there is no unique notion of entanglement entropy across the surface $S$.

In the following, we will exclusively consider pure spin network states, no linear combinations of spin networks. Then, if we additionally choose the recoupling scheme that diagonalizes the area operator for $S$ as the basis for the intertwiners at the punctures of $S$, the entanglement entropy becomes very simple. At a puncture $p$, with the intertwiner characterized by the quantum numbers 
$j^u$, $j^d$, $j^{u+d}$, the contribution to the entanglement entropies is
\begin{align}
\ent_\Omega(p)&=\ln(2j^d+1),\\
\ent_{\bar{\Omega}}(p)&=\ln(2j^u+1),\\
\ent_{S}(p)&= \ln(2j^{u+d}+1)
\end{align}
This result confirms \eqref{eq_nonunique}. However, there is a distinguished case where $\ent_\Omega = \ent_{\bar{\Omega}}$:
\begin{equation}
\ent_\Omega(p)=\ent_{\bar{\Omega}}(p) \quad \Longleftrightarrow \quad j^u=j^d. 
\end{equation}
In this case, we can meaningfully speak of \emph{the} entanglement entropy \emph{across} the surface $S$. Interestingly, in the case of the the three-leg vertex, comparison with the mean curvature operator of section \ref{sec_quantgeo} shows that these are precisely states for which the mean curvature of the boundary surface vanishes. In other words, the entanglement entropy across the surface $S$ is symmetric precisely when the surface is a minimal surface,  
\begin{equation}
\label{eq:minmal}
\ent_\Omega(p)=\ent_{\bar{\Omega}}(p) \quad \Longleftrightarrow \quad \widehat{H}(p)\Psi =0.  
\end{equation}
On the other hand, if the mean curvature of the boundary does not vanish, there will always be a jump $\triangle \ent(p) = \ent_{\Omega}(p)-\ent_{\bar{\Omega}}(p)$ in entanglement as one passes from one side of the surface $S$ to the other. 

\begin{figure}[!htbp]
\subfigure[$\jd=9$, $\ju=9, \, ...\, , 70$]{\includegraphics[width=.9\columnwidth,keepaspectratio]{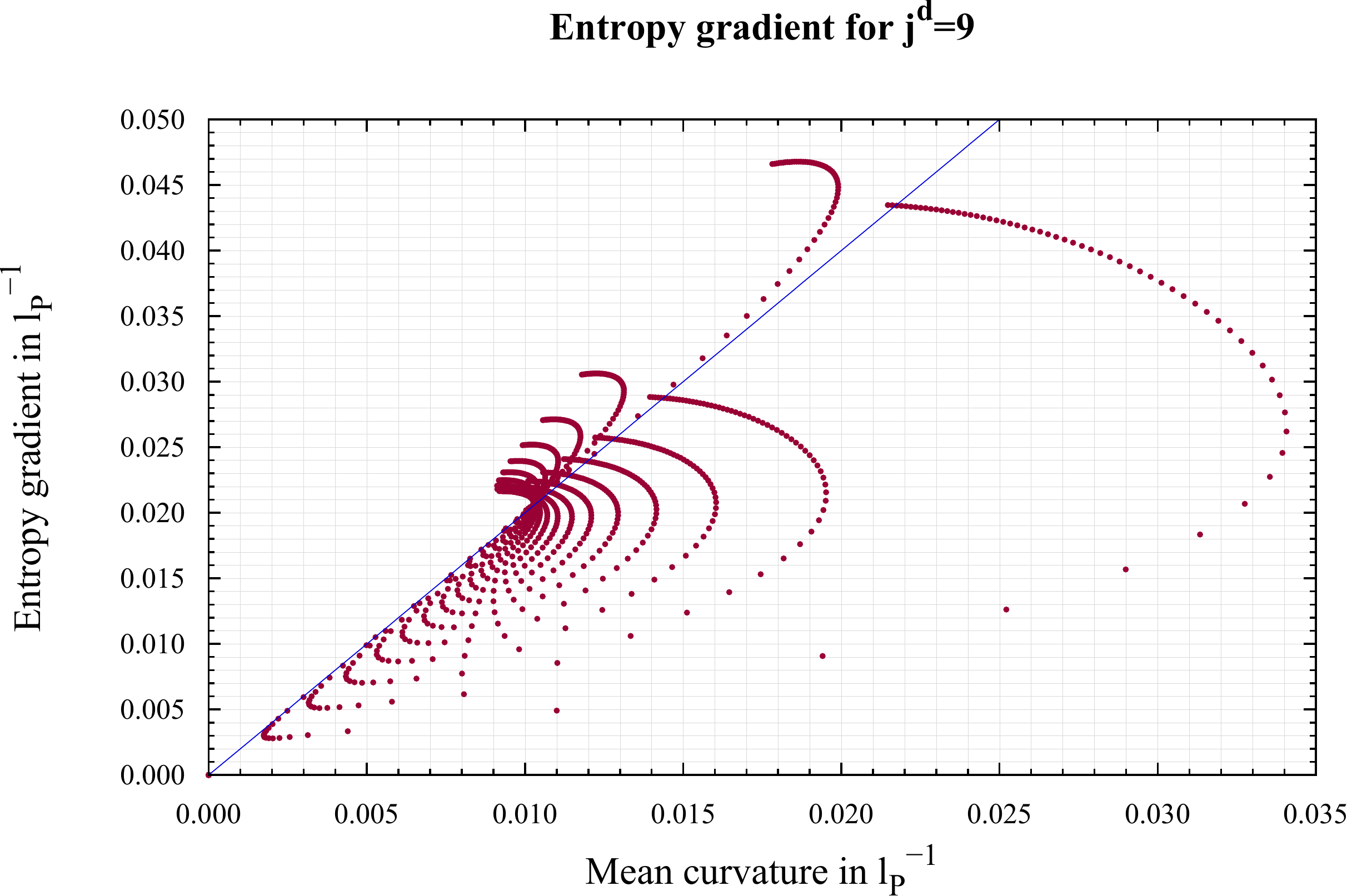}}
\subfigure[$\jd=60$, $\ju=60, \, ...\, , 80$]{\includegraphics[width=.9\columnwidth,keepaspectratio]{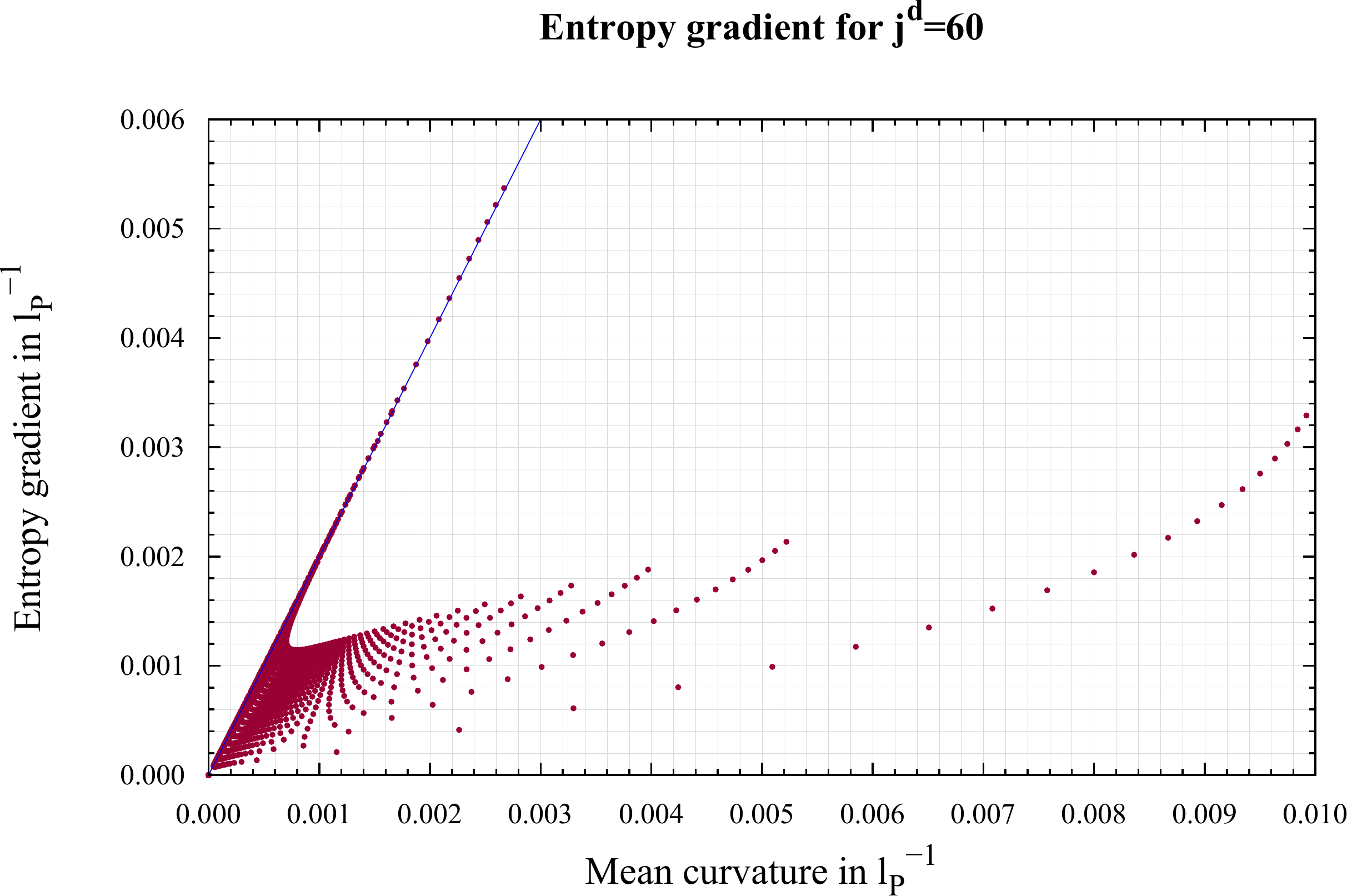}}
\caption{The discrete entropy gradient at the three-valent vertex is plotted against its mean curvature. For (a) $\jd=9$ is chosen and $\ju$ ranges from 9 to 70. In (b) we have set $\jd=60$ and $\ju=60, \, ...\, , 80$. There are different branches, corresponding to the possible choices of $\jud$. In blue we plotted a line through the origin with a slope of 2. One observes that for small mean curvatures there is a linear regime, whereas for larger values of $H$ the dependence gets highly non-linear.}
\label{fig:EntrvsMC}
\end{figure}

We want to examine the relation between $\ent$ and $H$ in the quantum theory in more detail. In particular, we want to argue that, for the three-leg vertex, there is a linear relationship between the gradient of $\ent$ across the surface and $H$. 
The classical expression \eqref{eq:extrinsicAreaDeri} for the mean curvature can be written as
\begin{equation}
H =\frac{1}{2A}\pd{A}{t} = \frac{1}{2} \pd{\ln{A}}{t} \,.
\end{equation}
The derivative is performed along an affinely parametrized geodesic in adapted coordinates. For large spins, the contribution to the area of the boundary surface from the vertex under investigation is proportional to the largest transversal spin: $A=a_0\cdot j$, where $a_0=4\pi\beta l_P^2$ and $j=\max{\{ \ju,\jd \}}$. The behavior is slightly different if $\jud$ becomes large compared to $j$. Hence, we obtain a proportionality between the mean curvature and the derivative of the entropy along a geodesic:
\begin{equation}
H\approx \frac{1}{2}\pd{\ln{\left( a_0\cdot j \right)}}{t} \sim \pd{\ln{\left( 2 j +1 \right) }}{t}= \pd{S}{t} \, .
\end{equation}
Note that the expressions contain continuous partial derivatives which a priori have no meaning in the quantum theory. Therefore, to be more precise, we will use a difference quotient below. But, for the moment, we want to imagine a semiclassical regime in which $j$ depends on a parameter $t$ that measures geodesic distance from $S$. Then, with $\kappa(t)\coloneqq 2j(t)+1$ and $\sigma\coloneqq j(t)(j(t)+1)$, we have
\begin{equation}
\begin{aligned}
H &= \frac{1}{2}\pd{\ln{\left( a_0\sqrt{\sigma} \right)}}{t} = \frac{1}{2}\pd{\ln{\left( \sqrt{ \sigma } \right)}}{t} \\
&= \frac{1}{4\sigma}\pd{\sigma}{t} = \frac{\kappa}{4\sigma}\pd{j}{t} \, ,
\end{aligned}
\label{eq:HvonJderi}
\end{equation}
and, on the other hand,
\begin{equation}
\pd{S}{t} = \frac{1}{\kappa} \pd{\kappa}{t} = \frac{2}{\kappa}\pd{j}{t} \, .
\label{eq:dsdt}
\end{equation} 
Comparison of \eqref{eq:HvonJderi} and \eqref{eq:dsdt} yields
\begin{equation}
H=\frac{\kappa^2}{8\sigma} \pd{S}{t} \coloneqq c_j  \pd{S}{t}. 
\label{eq:HasSderi}
\end{equation}
$c_j(t)$ depends only mildly on the spins involved, and 
\begin{equation}
\lim_{j \rightarrow \infty} c_{j} =\frac{1}{2}.
\end{equation}
But, as we remarked before, in the quantum theory there is no smooth $j(t)$. Let us rather consider a single puncture on $S$. We use a recoupling scheme such that $j^u$, $j^d$ and $j^{u+d}$ are well-defined. Then we have
\begin{align}
\triangle \ln a(p) &= \ln(a_0 \sqrt{j^u(j^u+1)})-\ln(a_0 \sqrt{j^d(j^d+1)})\\
&\approx \ln\left(\frac{j^u}{j^d}\right) \, ,
\end{align}
where $\triangle$ denotes the change that occurs when deforming $S$ slightly such that the vertex can lie above or below of $S$. The approximation is good for $j >> 1$. Similarly, we obtain
\begin{align}
\triangle S(p)&=\ent_{\bar{\Omega}}(p) - \ent_{\Omega}(p)=\ln(2j^u+1)-\ln(2j^d+1)\\
&\approx  \ln\left(\frac{j^u}{j^d}\right)\approx \triangle \ln a \, .  
\end{align}
Hence, at the puncture, the relation
\begin{equation}
\label{eq:discrete}
H(p)=\frac{1}{2} \frac{\triangle \ln a(p)}{\triangle t}\approx \frac{1}{2} \frac{\triangle S(p)}{\triangle t}
\end{equation}
holds. We have not yet specified $\triangle t$, but we can do so by using the length operator across the vertex. In figure \ref{fig:EntrvsMC}, we plotted the difference quotient\begin{equation}
\frac{\triangle S(p)}{\triangle t}=\frac{\ent_{\bar{\Omega}}(p)-\ent_{\Omega}(p)}{l^u(p)+l^d(p)}
\end{equation}
as a function of the mean curvature of the three-leg vertex. Here, the shift distances $l^u$ and $l^d$ where defined as in the case of the mean curvature operator: The lengths were measured along the holonomies transversal to the puncture $p$. 
Additionally, we plotted a line with a slope of 2 for comparison with equation \eqref{eq:HasSderi}. The plots in figure \ref{fig:EntrvsMC} show that, for small mean curvatures, the entropy gradient in fact shows a linear behavior. Small mean curvatures are obtained mainly if the transversal spins grow large while $\jud$ takes moderate values, away from its maximum and minimum. Moreover, we observe that there are many branches arising. They correspond to different values of the spin $\jud$. For larger mean curvatures, the branches leave the linear regime and bend towards larger entropy gradients, with slopes depending on $\jud$.\par 

\eqref{eq:HasSderi} and \eqref{eq:discrete} are generalizations of \eqref{eq:minmal}. The asymmetry in entanglement entropy $\ent(\Omega)-\ent(\bar{\Omega})$, as measured from above and below $S$ at a point $p$, is proportional to the mean curvature of the surface at $p$, at least in the three-leg case. This implies that, for minimal surfaces $S$ as boundaries, the entanglement entropy of the bulk spaces is conserved under variations of $S$.

The results we presented in this section are only valid for the three-leg puncture because the limit involved in the definition of the mean curvature operator \eqref{eq:defMeanCurvOp} implies that the areas in the formula are not simply determined by the coupled total upper/lower spin quantum numbers; the individual contributions of the edges above/below the surface directly influence the results. Thus, for example, we have
\begin{equation}
\lim_{\epsilon \rightarrow 0} \, \lim_{t \text{\tiny{$\searrow$}} 0} \, \widehat{A}(D(\epsilon,t))\neq 8\pi\beta l_{\text{P}}^2\sqrt{j^u(j^u+1)}\text{ id}
\end{equation}
in general. One could argue that a natural definition of this limit would be given by the right hand side of the previous equation. Using this definition -- and an analogous one for $j^d$ -- one would obtain a mean curvature operator for which the results of this section hold for arbitrary spin networks.

\section{Outlook} 
With the present work, we have explored the question whether we can assign intrinsic and extrinsic curvature to a surface $S$ in LQG. The results are mixed. 

One rather generic finding is that the scale characterizing the curvature spectra is the Planck scale. This means that the modulus of the curvatures assigned to the punctures by our method is typically extremely large. That makes it hard to assign any kind of classical picture to the surface at small scales. 
One notable class of states where the curvature can be small 
are the punctures where one transversal spin is large compared to the tangential spin. In this case, the scalar curvature $R$ tends to~$0$. A particular example of this is shown in figure~\ref{fig:intrinsic40-0}.
 
Moreover, we also generically obtain large coverage angles \eqref{eq:alphoperator}, see for example figure~\ref{fig:coverage80-60}. This is problematic regarding the entire definition of the intrinsic curvature operator which relies on formulas for circles that are small compared to the curvature radius.\\
We also found that the results are qualitatively the same if one bases the quantization on a picture in which the geometry near the punctures is cone-like.\\
As a result of the generically large curvatures, it does not seem that the integral over the scalar curvature in the quantum theory \eqref{eq:quantumgb} has much to do with the Euler characteristic of $S$. It is also a problematic sign that the curvature is not straightforwardly defined on generic points of $S$, i.e., in locations without punctures (see the discussion below \eqref{eq:KOperator} for details).\\ 
On the positive side, the limit in the classical formulas for intrinsic \eqref{eq:Rclass} and extrinsic curvature \eqref{eq:extrinsicAreaDeri} is well-defined, at least at vertices, also in the quantum theory, with the exception of some non-generic cases. 
Furthermore, we discovered, at least for a certain class of punctures, an intriguing connection between extrinsic curvature and entanglement entropy. Among other things, it implies that minimal surfaces enjoy a special property: the entropy gradient across them vanishes. This is interesting since many black hole horizons have slices that are spatial minimal surfaces. At least superficially, there also seems to be some hint of a connection to the results by Ryu and Takayanagi  \cite{Ryu:2006bv} and, more generally, to ER=EPR \cite{Maldacena:2013xja}, although in the former case, the entanglement entropy is that of a lower-dimensional theory, and it is given by the area of the minimal surface.\\

Our analysis is, among other things, limited by
\begin{itemize}
\item the use of kinematical states,
\item the ambiguities in the definition of the operators we defined, 
\item and the use of pure spin networks with three-leg punctures, as opposed to linear combinations of general spin networks like, for example, coherent states.  
\end{itemize}
Changing any of these might change the overall picture. Furthermore, there are different ways to approach the definition of curvature in the quantum theory. \cite{Alesci:2014aza} presents an example that can be adapted to lower-dimensional surfaces. Preliminary investigation indicates a better behavior in generic points of $S$, but more work would need to be done.  

\begin{acknowledgments}
TZ thanks the Elite Network of the State of Bavaria for financial support during the early stages of this work. The authors would like to thank the members of the Institute for Quantum Gravity at the Friedrich-Alexander-Universität Erlangen-Nürnberg for helpful discussions. 
\end{acknowledgments}

\bibliography{bibliography}

\end{document}